# Monoclinic phase in the relaxor-based piezo-/ ferroelectric Pb(Mg$_{1/3}$Nb$_{2/3}$)O$_3$ - PbTiO$_3$ system

Z.-G. Ye,[1] B. Noheda,[2] M. Dong,[1] D. Cox[2] and G. Shirane[2]

[1] Department of Chemistry, Simon Fraser University, Burnaby, BC, V5A 1S6, Canada

[2] Physics Department, Brookhaven National Laboratory, Upton, New York 11973 -500

A ferroelectric monoclinic phase of space group $Cm$ ($M_A$ type) has been discovered in 0.65Pb(Mg$_{1/3}$Nb$_{2/3}$)O$_3$–0.35PbTiO$_3$ by means of high resolution synchrotron X-ray diffraction. It appears at room temperature in a single crystal previously poled under an electric field of 43 kV/cm applied along the pseudocubic [001] direction, in the region of the phase diagram around the morphotropic phase boundary between the rhombohedral ($R3m$) and the tetragonal ($P4mm$) phases. The monoclinic phase has lattice parameters $a$ = 5.692 Å, $b$ = 5.679 Å, $c$ = 4.050 Å and $\beta$ = 90.15°, with the $b_m$-axis oriented along the pseudo-cubic [110] direction. It is similar to the monoclinic phase observed in PbZr$_{1-x}$Ti$_x$O$_3$, but different from that recently found in Pb(Zn$_{1/3}$Nb$_{2/3}$)O$_3$ - PbTiO$_3$, which is of space group $Pm$ ($M_C$ type).

PASC number(s): 77.84.Bw, 61.10.Nz

## I. INTRODUCTION

The complex perovskite Pb(Mg$_{1/3}$Nb$_{2/3}$)O$_3$ [PMN] exhibits typical relaxor-type ferroelectric properties which have been intensively studied for both fundamental and practical reasons.[1-3] The relaxor state is characterized by the frustration of local polarizations which prevents long-range ferroelectric order from developing completely. Although the local symmetry of the polar domains is rhombohedral, the macroscopic symmetry of PMN remains cubic below the temperature of maximum permittivity. A ferroelectric phase can be induced either by application of an electric field along [111]$_{cub}$,[4] or by partial substitution of Ti$^{4+}$ for the complex (Mg$_{1/3}$Nb$_{2/3}$)$^{4+}$ ions.[5] In both cases, the micropolar domains transform into macro domains with cubic-to-rhombohedral symmetry breaking.[4,6] With Ti-substitution, a complete solid solution forms between PMN and PbTiO$_3$, (1-x)PMN-xPT [PMN-PT], with a morphotropic phase boundary (MPB) located at x ≈ 0.35, which has hitherto been believed to represent the frontier between the rhombohedral (relaxor side) and tetragonal (ferroelectric side) phases,[5] as illustrated in Figure 1.

Single crystals of PMN–PT and other relaxor–PT solid solutions, namely Pb(Zn$_{1/3}$Nb$_{2/3}$)O$_3$ – PT [PZN-PT], with compositions near the respective MPB's, have been reported to exhibit very high piezoelectric coefficients (d$_{33}$ > 2,500 pC/N), extremely large piezoelectric strains (>1.7 %), and very high electromechanical coupling factors (k$_{33}$ > 92 %).[7-11] Such excellent properties point to a potential revolution in electromechanical transduction for a broad range of applications. From the crystal chemistry viewpoint, the strong piezoelectric response of those materials is associated with the MPB, and very recently, intensive theoretical and experimental research has been undertaken in order to understand the origin of these enhanced piezoelectric properties. The experimental breakthroughs have been achieved by the discovery of new ferroelectric monoclinic and orthorhombic phases by high-resolution synchrotron X-ray diffraction. In PbZr$_{0.52}$Ti$_{0.48}$O$_3$ [PZT48], a monoclinic phase with space group $Cm$ (of $M_A$ type, in the Vanderbilt and Cohen notation[12]) was found below room temperature. In such a phase, the monoclinic $b$-axis lies along pseudocubic [1$\bar{1}$0], and the polarization is in the (1$\bar{1}$0) plane.[13,14] In the PZN–PT system, an orthorhombic phase was found to exist between the rhombohedral and the tetragonal phases at around 10%PT.[15] Upon application of an electric field along [001]$_{cub}$, evidence of polarization rotation via a monoclinic phase has been shown for PZN-8%PT,[16] which is in agreement with the theoretical considerations by first principles calculations for the PZT case.[17, 18]

So far, several pictures of the phase states near the MPB of the PMN–PT system have been proposed. Shrout et al [5] initially suggested that the rhombohedral *R3m* phase of smaller Ti-concentration and the tetragonal *P4mm* phase of higher Ti-content, were separated by a fairly sharp phase boundary at 35%PT. Noblanc et al [19] reported a large composition range around MPB in which the rhombohedral and the tetragonal phases coexist. Optical domain studies showed that the *R3m* and *P4mm* phases may coexist at room temperature, with transformation from the *R3m* into the *P4mm* phase taking place over a wide temperature interval.[6] In the present paper, we report the results of a high-resolution synchrotron X-ray diffraction study of 0.65Pb(Mg$_{1/3}$Nb$_{2/3}$)O$_3$-0.35PbTiO$_3$ [PMN-35%PT] single crystals, showing unambiguous evidence of a monoclinic phase, which is stable between the rhombohedral and tetragonal phases near the MPB.

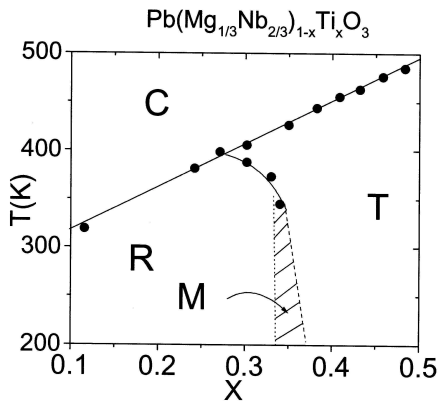

FIG. 1 Phase diagram of the PMN – PT solid solution system. Solid circles and the related phase boundaries are adapted from Ref. 5. *C*, *R* and *T* refer to cubic, rhombohedral and tetragonal regions, respectively. The diagonally shaded area represents the monoclinic *M* phase region, updated based on the present work.

## II. EXPERIMENTAL

Single crystals of PMN-35%PT were grown from a high temperature solution using an optimum flux composition (50wt%PMNT + 49wt%PbO + 1wt%B$_2$O$_3$).[20] Two platelets, #1, 65 µm thick, and #2, 165 µm thick, were cut parallel to the reference (001)$_{cub}$ plane and polished with fine diamond paste (down to 1 µm). The (001)$_{cub}$ faces were covered with Ag-paste, Au-wires were attached and the platelets #1 and #2 were poled under an electric field of 43 and 5 kV/cm, respectively, which was applied along [001]$_{cub}$ at 200 °C (above T$_C$) and maintained while cooling down to room temperature. The samples were then short-circuited for 30 min. before the electrodes were removed. Synchrotron X-ray diffraction experiments were performed at the Brookhaven National Synchrotron Light Source (NSLS). The PMN-35%PT crystal #1 poled at 43 kV/cm was studied at beam line X22A with 30 keV X-rays from the third order reflection of a Si(111) monochromator. The measurements were made on a Huber four-circle diffractometer equipped with a Si(111) analyzer. X-ray powder diffraction was also performed on PMN-35%PT at beam line X7A with 18 keV X-rays from a Si(111) double-crystal monochromator ($\lambda \approx 0.7$ Å). The PMN-PT crystal #2, previously poled at 5kV/cm, was crushed and sieved to produce a powder sample with large crystallites (38 – 44 µm), that we call a properly arranged single crystal (PASC), in order to obtain sharp powder lines free from sample-broadening effects. A 0.2 mm capillary loaded with the specimen was mounted on a Huber diffractometer equipped with a Ge(220) analyzer. The resulting instrumental resolution is about 0.005° on the 2θ scale.

## III. RESULTS AND DISCUSSION

Figure 2 gives three examples of the contour maps observed in different cubic zones around the (002), (300) and (330) reflection peaks for the PMN-35%PT crystal #1. For these experiments, the Bragg indices H=K=L=1 were defined to correspond to the cubic lattice parameters $a = b = c = 4.016$ Å for all the cases.

To interpret the diffraction patterns observed in Figure 2 and to illustrate the structure of the PMN-35%PT, we need to consider the possible domain configurations based on group/subgroup symmetry relationship. The high temperature prototype phase has a $Pm\bar{3}m$ space group symmetry. The phase transitions into a rhombohedral $R3m$ phase, or a monoclinic $m$ phase would normally result in eight or twenty-four (fully) ferroelectric / (partially) ferroelastic domain states, respectively.[21] The domain structures are therefore very complicated, especially for the monoclinic phase. However, a much-simplified situation is produced by electric field poling, which constrains the $c$-axis to lie along the pseudo-cubic [001] direction, analogous to the situation previously discussed for PZN-8%PT.[16] The monoclinic domain configuration now consists of two $b$-domains related by a 90°-rotation around the $c$-axis, each of which contains two domains.

The contour map of Fig. 2(a) shows a single d-spacing in the H0L zone measured around the (002) reflection, which gives the parameter $c = 4.050$ Å. The fact that a single d-spacing was observed in this region demonstrates that the $c$-axis is indeed fixed along the poling field direction of pseudocubic [001], and that the poled state remains stable after the field is removed. The very weak extra peaks observed in the transverse direction may correspond to slightly-misaligned regions close to the surface, since they were not observed with high-energy X-rays at 67 keV. Fig. 2(b) is the contour map obtained around the pseudocubic (300) peak in the H0L zone, and shows a symmetrical splitting along the transverse L direction. This splitting can be attributed to the two a-domains, but it is in sharp contrast to the three peaks observed for the (200) reflection from PZN-8%PT,[15] which has monoclinic symmetry of type $M_C$ (in the Vanderbilt and Cohen notation [12]). On the other hand, the contour map around the pseudocubic

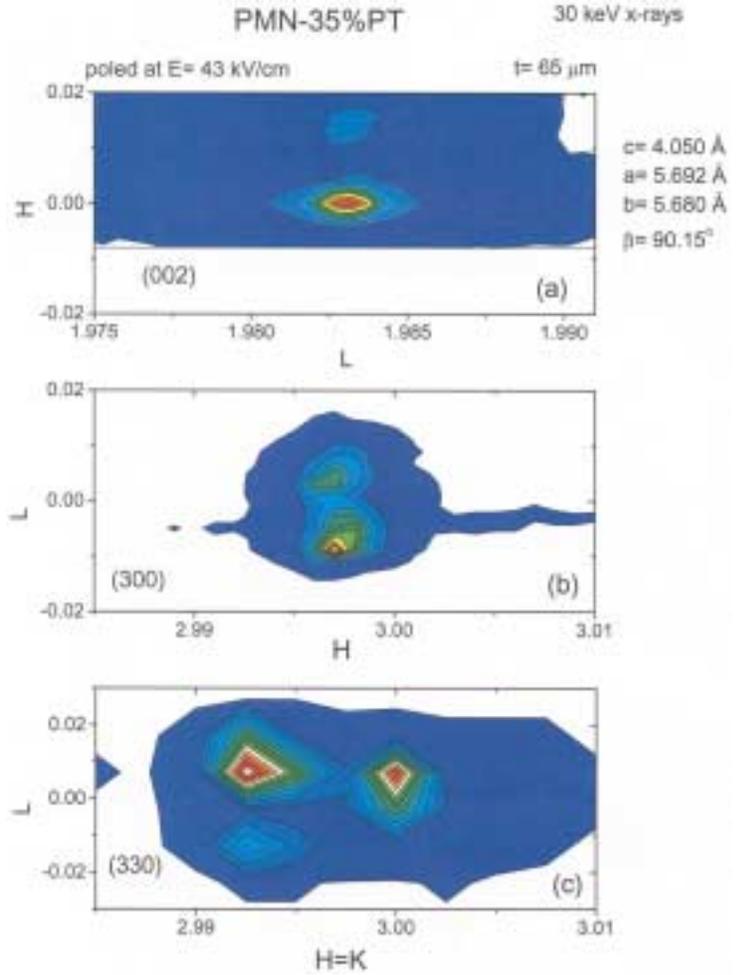

FIG. 2 Contour maps observed around (002), (300) and (330) reflection peaks in the cubic H0L (a), H0L (b) and HHL (c) zones, respectively, for a poled PMN-35%PT crystal. H=K=L=1 has been defined to correspond to d = 4.016 Å for all the cases.

(330) reflection shown in Fig. 2(c) is seen to contain three peaks, and in conjunction with Fig. 2(b) provides the signature of a monoclinic structure of $M_A$ type with $a_m$ and $b_m$ directed along pseudocubic $[\bar{1}\bar{1}0]$ and $[1\bar{1}0]$. The split peak is located at H = K = 2.993 and the single peak at H = K = 3.000, corresponding to lattice parameters $a_m = 5.692$ Å and $b_m = 5.679$ Å. The transverse separation of the split peaks is L = 0.022, yielding a monoclinic angle β = 90.15°. The projection of the monoclinic angle (β′ ≈ β√2) can also be deduced independently from the splitting of the (300) peak in Fig. 2(b), namely, L = 0.012, which yields a similar value of 90.16°. Finally, we note that although a rhombohedral domain structure could also give qualitatively similar features, such a structure can immediately be excluded, since the requirements of rhombohedral symmetry would impose constraints on the lattice parameters [*e.g.* $c_m^2 = (a_m^2 + b_m^2)/4$] which are clearly not satisfied. Thus we conclude that a monoclinic phase of $M_A$ type has been unambiguously shown to exist in the PMN-35%PT crystal. Figure 3 shows longitudinal and transverse linear scans for each of the (002), (300) and (330) reflections in Fig. 2. The lines are fits to Gaussian profiles. These scans confirm the monoclinic symmetry and lattice parameters deduced above.

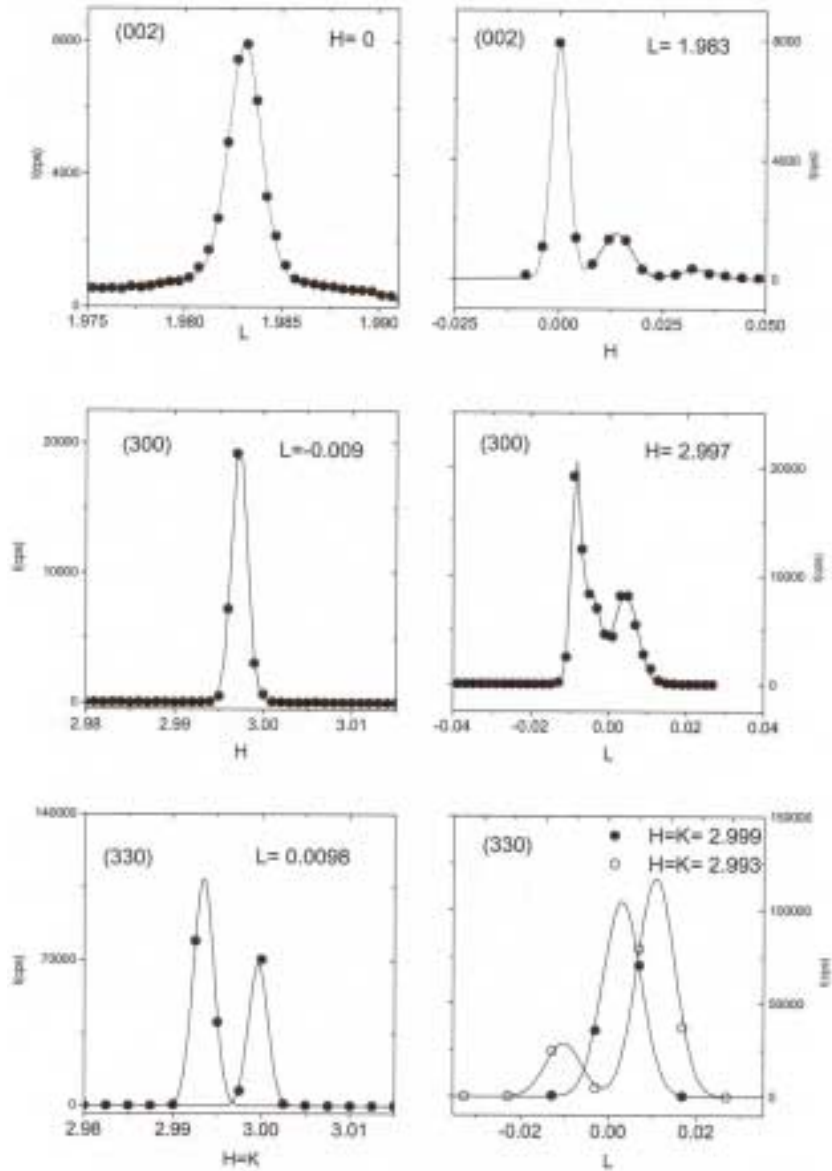

FIG. 3 Longitudinal and transverse linear scans for each of the (002), (300) and (330) reflections in Fig. 2. The lines are fits to Gaussian profiles.

We have also studied the weakly poled (E<5 kV/cm) PMN-35%PT crystal (#2) in the form of crushed (PASC) powders by X-ray powder diffraction at X7A. Figure 4 shows the sharp powder lines from a PASC sample with grain size between 38 and 44 μm. Only a single peak was observed for the $(100)_{cub}$ reflection, while both $(110)_{cub}$ and $(111)_{cub}$ reflections showed split doublets. These patterns can be clearly indexed as a single rhombohedral $R$ phase with $a = 4.023$ Å, $\alpha = 89.88°$. Therefore, unpoled or weakly poled PMN-35%PT samples show an average rhombohedral phase, consistent with the reported MPB phase diagram.[5]

## IV. CONCLUSIONS

A monoclinic phase of space group $Cm$ ($M_A$ type) has been discovered in a PMN-35%PT crystal at room temperature by means of high-resolution synchrotron X-ray diffraction. This $M_A$ phase has its monoclinic $a_m$- and $b_m$-axes oriented along pseudo-cubic $[1\bar{1}0]$ and $[110]$ directions, respectively, with the $c_m$-axis along pseudocubic $[001]$. The unit cell parameters are found to be $a_m$ ($\approx a_c \cdot \sqrt{2}$) = 5.692 Å, $b_m$ ($\approx a_c \cdot \sqrt{2}$) = 5.679 Å, $c_m$ ($\approx a_c$) = 4.050 Å and $\beta = 90.15°$. This monoclinic phase exhibits the same crystallographic features as the one reported for $PbZr_{1-x}Ti_xO_3$ (i.e. a doubled cell rotated by 45° around the $c$-axis with respect to the tetragonal unit cell),[13] but is different from the $M_C$–type monoclinic phase (space group $Pm$) recently reported in PZN-8%PT, where the monoclinic $b_m$-axis lies along $[0\bar{1}0]_{cub}$.[15]

The monoclinic phase appears at room temperature in a PMN-35%PT crystal previously poled under an electric field of 43 kV/cm applied along $[001]_{cub}$. On the other hand, unpoled and weakly-poled PMN-35%PT samples exhibit an average rhombohedral symmetry. Although poling at high-field along $[001]_{cub}$ may be necessary in order to produce a long-range monoclinic phase for the composition, this point is not yet fully settled and we hope to clarify it in future experiments. The existence of the monoclinic phase in PMN-35%PT allows us to construct an updated phase diagram for the PMN-PT solid solution system, in which a narrow monoclinic phase region exists between the rhomboheral $R3m$ and the tetragonal $P4mm$ phases at low temperatures, as indicated by the dashed area in Figure 1. More extended work is underway in order to provide a better understanding of the nature of the MPB and the origin of the outstanding piezoelectric performance of PMN-PT crystals, and will be reported elsewhere.

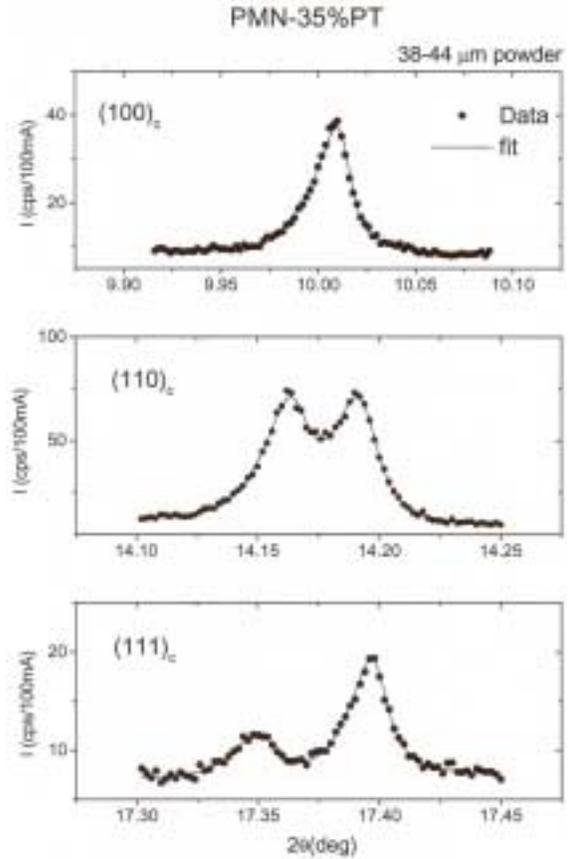

FIG. 4 $(100)_c$, $(110)_c$ and $(111)_c$ reflections obtained from powder diffraction of a properly arranged single crystal (PASC) sample made by carefully crushing a weakly poled (E<5 kV/cm) PMN-35%PT crystal.

After the submission of the manuscript, two important papers have appeared: Xu *et al.*[22] reported the observation of monoclinic domains by polarized light microscopy, and Lu *et al.*[23] showed the existence of an orthorhombic phase in 0.67PMN – 0.33PT crystals.

## ACKNOWLEDGMENTS

This work was supported by the ONR Grant # N00014-99-1-0738 and by the U.S. DOE under Contract No. DE-AC02-98CH10886. We thank John Hill, Ben Ocko and Zhong Zhong for very useful discussions.